\documentclass[a4paper, 10pt, 5p, preprint, twocolumn]{elsarticle}
\usepackage{amsmath,graphicx,color}
\usepackage[normalem]{ulem}
\usepackage[T1]{fontenc}
\usepackage{amssymb}
\usepackage{xcolor}
\usepackage{hyperref}
\usepackage{mathtools, cuted}

\usepackage{mwe}
\usepackage{placeins}

\journal{NPA 
}

\begin{document}

\begin{frontmatter}

\title{Transverse mass scaling of dilepton radiation off a quark-gluon plasma}

\author[irfu]{Maurice Coquet}
\cortext[mycorrespondingauthor]{Corresponding author}
\ead{maurice.louis.coquet@cern.ch}
\author[bielefeld]{Xiaojian Du}
\author[ipht]{Jean-Yves Ollitrault}
\author[bielefeld]{S\"oren Schlichting}
\author[irfu]{Michael Winn}

\address[irfu]{Universit\'e Paris-Saclay, Centre d'Etudes de Saclay (CEA), IRFU,  D\'epartement de Physique Nucl\'eaire (DPhN), Saclay, France}
\address[bielefeld]{Fakult\"at f\"ur Physik, Universit\"at Bielefeld, D-33615 Bielefeld, Germany}
\address[ipht]{Universit\'e Paris Saclay, CNRS, CEA, Institut de physique th\'eorique, 91191 Gif-sur-Yvette, France}

\begin{abstract}
The spectrum of dileptons produced by the quark-gluon plasma in an ultrarelativistic nucleus-nucleus collision depends only, to a good approximation, on the transverse mass $M_t$ of the dilepton.
This scaling is exact as long as transverse flow is negligible, and the system is in local thermal equilibrium.
We implement a state-of-the-art modelization of kinetic and chemical equilibration in the early stages of the evolution to study the modifications of the spectrum. 
Violations of $M_t$ scaling resulting from these effects are evaluated as a function of the shear viscosity to entropy ratio ($\eta/s$) that controls the equilibration time.
We determine the dependence of the spectrum on system size, centrality, rapidity, and collision energy.
We show that the quark-gluon plasma produces more dileptons than the Drell-Yan process up to invariant masses of order $M\sim 4$~GeV at LHC energies. 
Due to different kinematics, for a given $M_t$, the dependence of the dilepton yield on $M$ is opposite for the two processes, so that experiment alone can in principle determine which process dominates.    
\end{abstract}
\end{frontmatter}

\section{Introduction}\label{intro}

An ultrarelativistic nucleus-nucleus collision produces strongly-interacting matter which rapidly thermalizes into a hot quark-gluon plasma (QGP)~\cite{Busza:2018rrf}.
This plasma expands freely into the vacuum and eventually cools down into a gas of hadrons. 
Electron-positron and muon-antimuon pairs, referred to as dileptons, are created throughout the history of the QGP by quark-antiquark annihilation.
Once produced, they reach the detector without any further interaction, so that they probe the entire space-time dynamics, including the early stages of the collision.
In particular, they carry unique information about the thermalization of the QGP~\cite{Martinez:2008di}. 
Dileptons produced by the QGP can be separated from those produced later on in the hadronic phase using the invariant mass, $M$, of the pair as a selection criterion.
Specifically, the contribution of the QGP dominates for $M\gtrsim 1.2$~GeV~\cite{Rapp:2013nxa,Song:2018xca}.

This QGP dilepton production can be studied not only as a function of the invariant mass $M$, but also as a function of the momentum of the dilepton.
It has long been known~\cite{McLerran:1984ay} that at a given rapidity $y$, the spectrum $dN^{l^+l^-}/d^4K$, where $K$ is the 4-momentum of the dilepton, should depend only on the transverse mass $M_t\equiv \sqrt{M^2+k_t^2}$, where $k_t$ is the transverse momentum of the dilepton, provided that the QGP is in local equilibrium, and that the production occurs early enough that transverse flow is negligible. 
We improve over this seminal work by implementing a state-of-the-art treatment of pre-equilibrium dynamics~\cite{Coquet:2021lca}, still neglecting transverse flow. 
We study the effect of kinetic and chemical equilibration on the $M_t$ spectrum, and the deviations from $M_t$ scaling that they induce. 

In Sec.~\ref{s:scaling}, we explain when and why $M_t$ scaling is expected, and we rederive the expression obtained by McLerran and Toimela for the $M_t$ distribution~\cite{McLerran:1984ay}. 
In Sec.~\ref{s:dimensional}, we discuss qualitatively, using dimensional analysis, the effects of pre-equilibrium dynamics. 
Quantitative results are presented in Sec.~\ref{s:results}. 
The setup of our calculation is the same as in our previous work ~\cite{Coquet:2021lca}, in which we only calculated the mass spectrum of dileptons, integrated over momentum. 
We first briefly recall this setup, and we show our results for the QGP dilepton spectrum in Pb+Pb collisions at $\sqrt{s_{\rm NN}}=5.02$~TeV.
We show how the shear viscosity over entropy ratio $\eta/s$ at early times, which governs the thermalization of the QGP, can be extracted from the spectrum. 
In Sec.~\ref{s:backgrounds}, we evaluate the dilepton spectrum resulting from the Drell-Yan process, i.e, the annihilation of quarks and antiquarks belonging to incoming nuclei~\cite{Drell:1970wh}, and we compare this background with the QGP spectrum. 


\section{Transverse mass scaling}
\label{s:scaling}

It has long been known that in a hadronic gas in thermal equilibrium, the transverse momentum spectra of all identified hadrons fall on the same curve when plotted as a function of their transverse mass $M_t$~\cite{Hagedorn:1965st,Aachen-Berlin-Bonn-CERN-Cracow-Heidelberg-Warsaw:1976gom,Becattini:1997rv,Altenkamper:2017qot}. 
This follows from the fact that the phase-space distribution of particles within the gas is a Boltzmann factor $dN/d^3p d^3x=\exp(-E/T)$, where $E$ is the energy of the particle, $T$ the temperature, and we have neglected the small effects of quantum statistics.
Writing $E=M_t\cosh y$, where $y$ is the rapidity, and integrating over the rapidity, the resulting distribution only depends on $M_t$.

This argument does not immediately apply to dileptons because they cease to interact as soon as they are produced.
Therefore, dileptons produced by an equilibrated QGP are not themselves in thermal equilibrium. 
However, transverse mass scaling still holds, for reasons which we now explain.

Dilepton production occurs through the production of a virtual photon, which then decays into a lepton-antilepton pair.
What one calls the 4-momentum of the dilepton, $K$, is actually the 4-momentum of the virtual photon.
To leading order in perturbation theory, the production of the virtual photon occurs through a $2\to 1$ process:
The annihilation of a quark, with 4-momentum $P_1$, and an antiquark, with 4-momentum $P_2$, into a virtual photon with 4-momentum $K=P_1+P_2$. 
We denote the phase space distributions of quarks and antiquarks by $f_q(x,{\bf p}_1)$ and $f_{\bar q}(x,{\bf p}_2)$, respectively, where $x$ denotes space-time coordinates.
The rate of dilepton production is obtained by integrating the transition rate over all possible values of ${\bf p}_1$ and ${\bf p}_2$, taking energy-momentum conservation into account: 

\begin{strip}
\begin{equation}
\label{prodrate}
  \frac{dN^{l^+l^-}}{d^4xd^4K}=\int{\frac{d^3 p_1}{(2\pi)^3 2p_1}\frac{d^3 p_2}{(2\pi)^3 2p_2}f_q(x,{\bf p}_1)f_{\bar q}(x,{\bf p}_2)|{\cal A}|^2 (2\pi)^4\delta^{(4)}(P_1+P_2-K)},
\end{equation}
where ${\cal A}$ is the Lorentz-invariant amplitude of the process, which is not modified by the thermal medium to leading order in perturbation theory. 
We have neglected quark masses so that the invariant phase-space element is just $d^3p_i/p_i$, with $i=1,2$, and $p_i\equiv |{\bf p}_i|$.
In a baryonless thermal fluid at rest, $f_q(x,{\bf p})=f_{\bar q}(x,{\bf p})=\exp(-p/T(x))$ (assuming Boltzmann statistics), where $T(x)$ is the temperature at space-time point $x$. 
Conservation of energy then implies $f_q(x,{\bf p}_1)f_{\bar q}(x,{\bf p}_2)=\exp(-k_0/T(x))$, which can be moved outside the integral: 
\begin{equation}
\label{prodrate2}
  \frac{dN^{l^+l^-}}{d^4xd^4K}=e^{-k_0/T(x)}\int{\frac{d^3 p_1}{(2\pi)^3 2p_1}\frac{d^3 p_2}{(2\pi)^3 2p_2}|{\cal A}|^2 (2\pi)^4\delta^{(4)}(P_1+P_2-K)}.
\end{equation}
\end{strip}

The pre-factor on the right-hand side is the Boltzmann factor corresponding to the dilepton, and the rest is a non-trivial kinematic integral involving the scattering amplitude, which could in principle depend on the four-momentum $K$. 
We show that it is in fact independent of $K$. 

First, one notes that the integrand is a Lorentz scalar. 
If one neglects quark and lepton masses, the only Lorentz-invariant scale is the invariant mass $M=(K^\mu K_\mu)^{1/2}$ of the dilepton, hence the integral can only depend on $M$. 
This dependence can be obtained through dimensional analysis. 
The left-hand side of Eq.~(\ref{prodrate2}) is dimensionless in natural units $\hbar=c=1$, therefore, the integral in the right-hand side is also dimensionless.
This implies that it is actually independent of $M$. 

For a fluid at rest, we have demonstrated that 
\begin{equation}
\label{prodrateC}
\frac{dN^{l^+l^-}}{d^4xd^4K}=C\exp\left(-\frac{k_0}{T(x)}\right),
\end{equation}
where $C$ is a dimensionless constant. 
Now, since $dN^{l^+l^-}/d^4xd^4K$ is a Lorentz scalar, the result for a moving fluid is identical, provided that one replaces  $k_0$ with the dilepton energy in the rest frame of the fluid.

The dilepton spectrum is obtained by integrating the production rate over the space-time coordinates $x^\mu$. 
We assume that the QGP is invariant under longitudinal boosts~\cite{Bjorken:1982qr}.
Then, its space-time volume can be rewritten as $d^4x=d^2{\bf x}_\perp  \tau d\tau dy_f$, where ${\bf x}_\perp$ is the transverse position, $\tau\equiv\sqrt{t^2-z^2}$ the proper time and $y_f={\rm artanh}(z/t)$ the fluid rapidity. 
Finally, we neglect transverse flow.
Then, the dilepton energy in the fluid rest frame is $M_t\cosh(y-y_f)$.
The dilepton spectrum is:
\begin{equation}
  \label{mtscalinggeneral}
  \frac{dN^{l^+l^-}}{d^4K}=C\int d{\bf x}_\perp \int_0^{\infty} \tau d\tau \int_{-\infty}^{+\infty} dy_f \exp\left(-\frac{M_t\cosh(y- y_f)}{T({\bf x}_{\perp},\tau)}\right). 
\end{equation}
It is independent of $y$, as a consequence of the assumed longitudinal boost invariance.
It depends on $k_t$ and $M$ only through $M_t$, which is the property of transverse mass scaling. 
As can be seen from the above argument, $M_t$ scaling is a robust property of the leading-order dilepton production, which is independent of the detailed space-time dynamics.  As long as the system is longitudinally boost invariant and the transverse expansion can be neglected it simply follows from symmetry and dimensional analysis. 
One expects it to be broken by next-to-leading order perturbative corrections, which are smaller than the leading-order contribution~\cite{Ghiglieri:2014kma}, and by non-perturbative dynamics~\cite{Ding:2016hua,Jackson:2019yao}. 
For leading-order production, an explicit calculation gives the expression of the proportionality constant $C$ in Eq.~(\ref{prodrateC})~\cite{Coquet:2021lca,Laine:2015iia}:
\begin{equation}
\label{expressionC}
C=\frac{N_c\alpha^2}{12\pi^4}\sum_f q_f^2,
\end{equation}
where $N_c=3$ is the number of quark colors, $\alpha$ is the fine structure constant, $\sum_f$ denotes the summation over quark flavors, $q_f$ is the quark electric charge, $\frac{2}{3}$ for $u$ and $-\frac{1}{3}$ for $d$ and $s$. 

We now derive the explicit form of the $M_t$ spectrum assuming that the equation of state of the QGP is conformal~\cite{Baier:2007ix}, which is approximately true at high temperatures, and implies that $\tau T({\bf x}_\perp,\tau)^3$ is independent of $\tau$~\cite{Bjorken:1982qr}. 
Then, the integral over $\tau$ in Eq.~(\ref{mtscalinggeneral}) can easily be done analytically using the change of variables $\tau=x^3$.
The integral over the fluid rapidity $y_f$ can also be done analytically. 
We further simplify the problem (although this simplification is not essential) by assuming that the temperature profile is uniform within a transverse area $A_\perp$, that is, $T({\bf x}_\perp,\tau)$ is independent of ${\bf x}_{\perp}$.
We obtain:
\begin{equation}
\label{ratethermal3}
\left(\frac{dN^{l^{+}l^{-}}}{d^4 K}\right)_{\rm ideal}=\frac{32N_c\alpha^2\sum_f q_f^2}{\pi^4} \frac{A_\perp (\tau T^3)^2 }{M_t^6},
\end{equation}
which is the McLerran-Toimela $M_t^{-6}$ spectrum~\cite{McLerran:1984ay}. 
The subscript {\it ideal\/} refers to the fact that local equilibrium holds at all times, which in turn implies that the expansion is ruled by {\it ideal\/} hydrodynamics. 

Note that the $M_t$ spectrum is a power law. 
This seems to contradict the expectation from lower energies that the $M_t$ spectrum should be exponential, with the inverse slope measuring the effective temperature probed by dileptons~\cite{NA60:2008ctj,Rapp:2014hha,Tripolt:2020dac,HADES:2019auv}. 
The contradiction is only apparent. 
Larger values of $M_t$ are produced at earlier times, when the temperature is higher. 
Therefore, the effective temperature depends on $M_t$: 
\begin{equation}
\label{teffideal}
T_{\rm eff}(M_t)\equiv -\left[\frac{d}{dM_t}
\ln\left(\frac{dN^{l^{+}l^{-}}}{d^4 K}\right)\right]^{-1}=\frac{M_t}{6}.
\end{equation}
It is the integration over time which converts the exponential spectrum into a power law~\cite{McLerran:1984ay}.

The constant $\tau T^3$ in Eq.~(\ref{ratethermal3}) is proportional to the charged multiplicity per unit rapidity~\cite{Kajantie:1986cu}, and inversely proportional to $A_\perp$~\cite{Coquet:2021lca}. 
Our estimates for Pb+Pb collisions at $\sqrt{s_{\rm NN}}=5.02$~TeV near mid-rapidity in the $0-5\%$ centrality range are:  
\begin{eqnarray}
\label{tauT3}
A_\perp&=&104~{\rm fm}^{2}=2670~{\rm GeV}^{-2}\cr
\tau T^3&=&10.3~{\rm fm}^{-2}=0.40~{\rm GeV}^2. 
\end{eqnarray}
These values will be used in Sec.~\ref{s:results}, where we show that our numerical results smoothly converge to Eq.~(\ref{ratethermal3}) when the viscosity over entropy ratio $\eta/s$, which controls the deviations from equilibrium, goes to zero. 

Finally, the spectrum (\ref{ratethermal3}) can be integrated over $M_t$ for fixed $M$. The resulting invariant mass spectrum is proportional to $M^{-3}$~\cite{McLerran:1984ay}:
\begin{eqnarray}
\label{rateintegrated}
  \left(\frac{dN^{l^{+}l^{-}}}{dMdy}\right)_{\rm ideal}&= &2\pi M \int_{M}^{\infty}M_tdM_t
  \left(\frac{dN^{l^{+}l^{-}}}{d^4 K}\right)_{\rm ideal}\cr
&=&  \frac{16N_c\alpha^2\sum_f q_f^2}{\pi^3} \frac{A_\perp (\tau T^3)^2}{M^3}.
\end{eqnarray}

\section{Pre-equilibrium dynamics: qualitative discussion}
\label{s:dimensional}

We now discuss qualitatively the effects of pre-equilibrium dynamics. 
Several effects must be taken into account: 
\begin{itemize}
    \item The time dependence of the temperature is modified due to the anisotropy of the momentum distribution of quarks and gluons. 
    \item The quark momentum distribution entering the production rate (\ref{prodrate}) is anisotropic. 
    \item Quarks are underpopulated relative to gluons.
\end{itemize}
The first two effects correspond to kinetic equilibration, while the third corresponds to chemical equilibration. 
At the end of this section, we also discuss the qualitative effects of transverse flow, which is not included in our numerical results. 

For this qualitative discussion, we model the departure from local thermal equilibrium by replacing ideal hydrodynamics with Navier-Stokes viscous hydrodynamics. 
The relative order of magnitude of viscous corrections can then be derived on the basis of dimensional analysis. 
The largest term in the energy-momentum tensor of an ideal fluid is proportional to $\epsilon+P=Ts$~\cite{Ollitrault:2007du}. 
The correction involving the shear viscosity $\eta$ is a gradient~\cite{Baier:2007ix}.
In the early stages of the collision, due to the fast longitudinal expansion, the largest gradient is the time derivative, which is of order $1/\tau$ for dimensional reasons.
Hence, the viscous term is of order $\eta/\tau$, while the ideal term is of order $Ts$.
The ratio of the two is the inverse Reynolds number, which depends on $\tau$:  
\begin{equation}
  \label{reynolds}
{Re}^{-1}(\tau) \equiv \frac{\eta}{s}\frac{1}{\tau T(\tau)}. 
\end{equation}
Now, dileptons with a transverse mass $M_t$ are dominantly produced when the temperature $T(\tau)$ is of the order of $M_t$.
This occurs at a time $\tau$ proportional to $(\tau T^3)/M_t^3$, where we recall that $\tau T^3$ is approximately constant. 
Inserting these orders of magnitude of $T(\tau)$ and $\tau$ into Eq.~(\ref{reynolds}), we obtain the order of magnitude of the  relevant Reynolds number, which now depends on $M_t$:
 \begin{equation}
  \label{reynoldsmt}
{Re}^{-1}(M_t) \equiv  \frac{\eta}{s}\frac{M_t^2}{\tau T^3}. 
\end{equation}
The relative correction to the dilepton yield due to pre-equilibrium dynamics is of the order of ${Re}^{-1}(M_t)$. 
As we shall see in Sec.~\ref{s:results}, this dimensional reasoning is confirmed by numerical calculations.
 
We now discuss, still at the qualitative level, the breaking of $M_t$ scaling which is expected when the plasma is not in local equilibrium. 
The first effect is that the quark momentum distribution is no longer isotropic.
Due to the fast longitudinal expansion, longitudinal momenta in the comoving frame are typically much smaller than tranverse momenta~\cite{Martinez:2008di}. 
In order to evaluate the qualitative effect of this momentum anisotropy on dilepton emission, we consider the extreme case where quark distributions are purely transverse, still assuming, for simplicity, that they are Boltzmann distributions: 
\begin{equation}
\label{2dboltzmann}
  f_{q,\bar q}(x,{\bf p})\propto \delta(p_{z})\exp(-p/T(x)), 
\end{equation}
where the proportionality factor has dimension of energy. 
Inserting this expression into Eq.~(\ref{prodrate}), and dropping the constant proportionality factors, one obtains 
\begin{equation}
\label{rate2d}
  \frac{dN^{l^+l^-}}{d^4xd^4K}\propto e^{-k^0/T(x)} \delta(k_z)\int{\frac{d^2 p_1}{p_1}\frac{d^2 p_2}{p_2}|{\cal A}|^2 \delta^{(3)}(P_1+P_2-K)},
\end{equation}
where the integration only runs over the transverse momenta, and we have factored out $\delta(k_z)$, so that the Dirac constraint inside the integral is now in 2+1 dimensions (transverse momentum and energy). 
The integrand is invariant under Lorentz transformations in 2+1 dimensions, which again implies that the integral can only depend on the invariant mass of the dilepton, $M$. 
Dimensional analysis of Eq.~(\ref{prodrate}) shows that the scattering amplitude ${\cal A}$ is dimensionless, such that the integral in Eq.~(\ref{rate2d}) has the mass dimension $-1$, and is therefore proportional to 1/M. 
Explicitly, the integral evaluates to $2 \pi |{\cal A}|^2/M$. 
The factor $\delta(k_z)$ can be rewritten as $(1/M_t)\delta(y)$, where $y$ is the rapidity of the dilepton. 
In a reference frame where the fluid has rapidity $y_f$, this becomes $(1/M_t)\delta(y-y_f)$.
Finally, since $k_z=0$, the energy of the dilepton coincides with its transverse mass, and one obtains:
\begin{equation}
\label{rate2d2}
\frac{dN^{l^{+}l^{-}}}{d^4x d^4 K}\propto \exp\left(-\frac{M_t}{T(x)}\right)\frac{1}{M_t}\delta(y-y_f) \frac{1}{M}.
\end{equation}
The dilepton spectrum is obtained by integrating over the space-time history of the fluid, as in Eq.~(\ref{mtscalinggeneral}). One obtains
\begin{equation}
  \label{2dspectrum}
  \frac{dN^{l^+l^-}}{d^4K}\propto \frac{1}{M_t M}\int d{\bf x}_\perp \int_0^{\infty} \tau d\tau \exp\left(-\frac{M_t}{T({\bf x}_{\perp},\tau)}\right). 
\end{equation}
$M_t$ scaling is broken by the factor $1/M$, which results from the reduced dimensionality of the phase-space integral in Eq.~(\ref{rate2d}). 
For a given $M_t$, $dN^{l^+l^-}/d^4K$ is smaller for larger values of $M$. 
As we shall see in Sec.~\ref{s:results}, this hierarchy is borne out by numerical calculations. 

The other effect of pre-equilibrium dynamics is that the relative abundances of quarks and antiquarks are smaller than thermal abundances in the early stages of the collision.
The collision between the incoming nuclei creates mostly gluons~\cite{Shuryak:1992wc}.
Quark-antiquark pairs are then gradually produced by collisions between gluons~\cite{Kurkela:2018oqw,Du:2020dvp}.
Since dileptons are produced by quark-antiquark annihilation, quark suppression implies a suppression of dilepton production~\cite{Shuryak:1992bt}. 

Studies of QCD thermalization in kinetic theory~\cite{Du:2020dvp} suggest that kinetic and chemical equilibration are governed by a single time scale, $\tilde{w}=\tau T /(4\pi\eta/s)$ , which can be seen as the age of the system $\tau$ in units of the equilibrium relaxation time $\sim (4\pi\eta/s)/T$. Hence, within our effective description the shear viscosity over entropy ratio $\eta/s$ controls not only the kinetic equilibration but also the approach to chemical equilibration.
Therefore, the suppression of dilepton production due to quark suppression follows the above dimensional analysis, and should scale like $Re^{-1}(M_t)$ in Eq.~(\ref{reynoldsmt}). 
It depends only on $M_t$, so that quark suppression by itself should not break $M_t$ scaling. 
However, it occurs in the early stages where the largest breaking of $M_t$ scaling is expected.
Therefore, one expects the breaking to be {\it milder\/} when quark suppression is included. 
We will check this in Sec.~\ref{s:results}. 

The last effect which breaks $M_t$ scaling is transverse flow. 
The transverse fluid velocity is proportional to $\tau$ at early times~\cite{Ollitrault:2007du,Vredevoogd:2008id,Kurkela:2018wud}.
Therefore, transverse flow becomes more and more important as time goes by.
Since the time of dilepton production decreases with $M_t$ like $M_t^{-3}$, one expects that effects of transverse flow become negligible if $M_t$ is large enough~\cite{Rapp:2014hha}. 
The qualitative effect of transverse flow on $M_t$ scaling is the following~\cite{Deng:2010pq}:
For a given $M_t$, the transverse boost enhances dilepton production for larger $k_t$ or, equivalently, smaller values of $M$. 
Note that this effect is qualitatively similar to the effect of pre-equilibrium dynamics discussed above. 
It has been seen experimentally by the NA60 Collaboration~\cite{NA60:2007lzy}. 
We do not model transverse flow, therefore, we cannot assess quantitatively the breaking of $M_t$ scaling resulting from it.
However, we will estimate in Sec.~\ref{s:results} the range of $M_t$ for which transverse flow is likely to be important.

\section{Pre-equilibrium dynamics: quantitative results}
\label{s:results}

We now present quantitative estimates of QGP dilepton production in Pb+Pb collisions at $\sqrt{s_{\rm NN}}=5.02$~TeV.
The calculation is essentially the same as in Ref.~\cite{Coquet:2021lca}, therefore we only recall the essential steps.
Compared to the calculation of Sec.~\ref{s:scaling}, the main difference lies in the quark momentum distribution $f_{q,\bar q}({\bf p})$ in Eq.~(\ref{prodrate}).
The momentum anisotropy is modeled by carrying out the following replacement~\cite{Martinez:2008di} in the Boltzmann\footnote{The only difference with Ref.~\cite{Coquet:2021lca} is that we use Boltzmann distributions instead of  Fermi-Dirac distributions in Eq.~(\ref{prodrate}). The advantage of this simplification is that our results converge to the McLerran-Toimela spectrum (\ref{ratethermal3}) as $\eta/s\to 0$, which is a useful benchmark.  We have checked that the dilepton yields decrease only by a few percent if one uses Fermi-Dirac instead of Boltzmann.} distribution:
\begin{equation}
  \label{xi}
|{\bf p}|\rightarrow \sqrt{p_t^2+\xi^2p_z^2},
\end{equation}
where $\xi>1$ is the anisotropy parameter.
Note that this ansatz implicitly assumes that the tail of the momentum distribution is exponential.
Therefore, our modelization does not take into account the possibility that the falloff at large momentum is slower than exponential, corresponding to the presence of an increased number of high-momentum partons in the early stages, usually referred to as ``minijets'' ~\cite{Paatelainen:2013eea}.
Note that, on the other hand, some choices of initial conditions inspired by the color glass picture imply a falloff at large momentum which is faster than exponential~\cite{Churchill:2020uvk}.

We take quark suppression into account by multiplying the Boltzmann distribution by a global ``quark suppression'' factor $q_s$, which is smaller than unity. 
The anisotropy parameter $\xi$ and the quark suppression factor are computed as a function of time using QCD kinetic theory~\cite{Du:2020dvp}. 
More precisely, we use QCD kinetic theory to evaluate the pressure anisotropy and the fraction of energy density carried by quarks.
We then match the anisotropy parameter and the quark suppression factor to these results.
Note that we could have used QCD kinetic theory to calculate directly the quark distribution.
The reason why we choose not to do so is the following.
There is by now strong theoretical evidence that the evolution of the pressure anisotropy is fairly universal~\cite{Heller:2015dha,Heller:2016rtz} and does not depend on the details of the microscopic dynamics~\cite{Giacalone:2019ldn}.
We therefore believe that the results we obtain in this indirect way, through a minimal distortion of the Boltzmann distribution,  provide an efficient, transparent and robust way to investigate the dilepton spectrum.
We recall that both out-of-equilibrium parameters, namely the anisotropy parameter $\xi$ and quark suppression $q_s$ depend on the scaling variable $\tilde{w}=\tau T /(4\pi\eta/s)$ as described in detail in \cite{Coquet:2021lca}. Hence the only free parameter in the calculation is the viscosity over entropy ratio $\eta/s$, which is assumed to be constant and controls the proper-time dependence of quark production and kinetic equilibration, e.g. for larger values of $\eta/s$, the quark distribution will approach isotropy and chemical equilibrium later on ($\tau\sim3 $fm/c for $\eta/s=0.32$) than for lower values ($\tau\sim1$fm/c for $\eta/s = 0.16$).

The other difference with the calculation of Sec.~\ref{s:scaling} is that the temperature decreases more slowly than $\tau^{-1/3}$ at early times, due to the smaller longitudinal pressure.
One recovers the $\tau^{-1/3}$ dependence at late times. 
The temperature is determined by matching the value of $\tau T^3$ at late times to the observed multiplicity. 
That is, the value of $\tau T^3$ at late times is the same as in the ideal case. 

Our numerical calculations are carried out for Pb+Pb collisions at $\sqrt{s_{\rm NN}}=5.02$~TeV in the  0-5\% centrality range, and the corresponding normalizations are given by Eq.~(\ref{tauT3}). 
We have carried out calculations with and without quark suppression, for four different values of $\eta/s$: $0.04$, $0.08$ (not shown), $0.16$, and $0.32$. 
The expected value for QCD, in the temperature range spanned by the early evolution, typically lies between $0.16$ and $0.32$~\cite{Christiansen:2014ypa}. 
Smaller values $0.04$ and $0.08\simeq\frac{1}{4\pi}$~\cite{Policastro:2001yc} have also been implemented, in order to check numerically that our results converge smoothly to McLerran-Toimela spectrum (\ref{ratethermal3}) in the limit $\eta/s\to 0$. 

\begin{figure}[ht]
\begin{center}
\includegraphics[width=\linewidth]{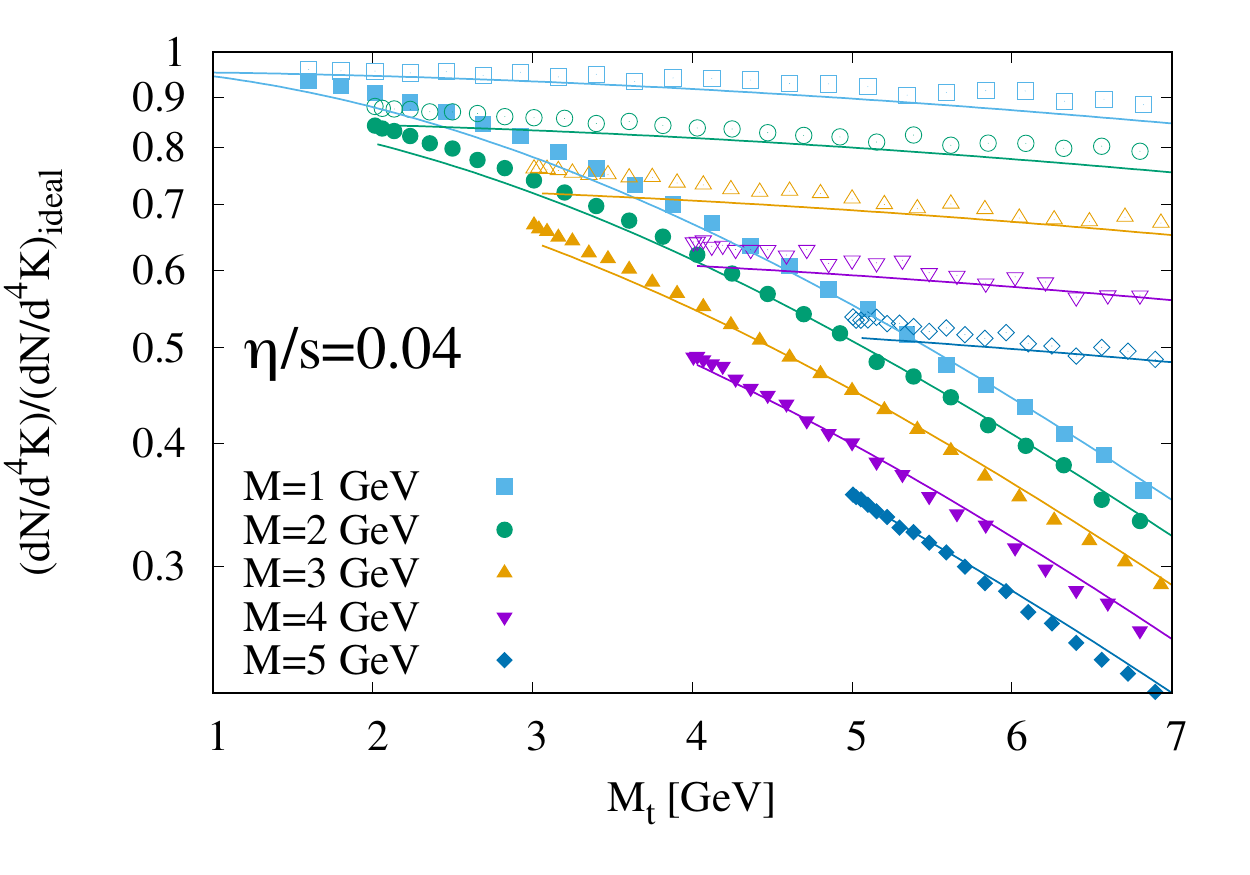} 
\end{center}
\caption{(Color online)
Ratio of the dilepton spectrum, evaluated with $\eta/s=0.04$, to the McLerran-Toimela spectrum (\ref{ratethermal3}), as a function of the transverse mass of the dilepton $M_t$, for several values of the invariant mass $M$. 
Calculations are for $0-5\%$ central Pb+Pb collisions at $\sqrt{s_{\rm NN}}=5.02$~TeV. 
Closed symbols correspond to our numerical calculations with all pre-equilibrium effects taken into account. 
Open symbols correspond to calculations where chemical equilibrium is assumed at all times, i.e., quark suppression is not taken into account. 
The thin lines are the global fit of our results using Eq.~(\ref{fitformula}) (see text). 
}
\label{fig:004results}
\end{figure}    
The results with $\eta/s=0.04$ are displayed in Fig.~\ref{fig:004results} for five equally-spaced values of the invariant mass $M$.\footnote{The lowest value $M=1$~GeV is shown only for the sake of illustration, as hadronic production, which we do not consider, is significant for $M<1.2$~GeV.}
We have divided the spectrum calculated numerically with the analytic result for ideal hydrodynamics, Eq.(\ref{ratethermal3}). 
The ratio is smaller than unity, which confirms the expectation that pre-equilibrium effects inhibit dilepton emission. 
It is naturally smaller when quark suppression is taken into account, as can be seen by comparing closed symbols with open symbols. 
The ratio is very close to unity for small $M_t$.
The deviation from unity increases as a function of $M_t$ as expected from the larger deviations from equilibrium at the time of production in Eq.~(\ref{reynoldsmt}). 
For a fixed $M_t$, the yield decreases as the invariant mass $M$ increases, in line with the expectation from Eq.~(\ref{2dspectrum}).  

The dependence of the dilepton yield on the parameters $\eta/s$, $M_t$ and $M$ is well captured by the following formula: 
\begin{equation}
\label{fitformula}
\frac{dN^{l^{+}l^{-}}}{d^4 K}\simeq
\left(\frac{dN^{l^{+}l^{-}}}{d^4 K}\right)_{\rm ideal}
\frac{\left(1+a\frac{\displaystyle\eta}{\displaystyle s}M_t^2/n\right)^{-n}}{\sqrt{1+b\frac{\displaystyle\eta}{\displaystyle s}M^2}} 
\end{equation}
where the first term in the right-hand side is the McLerran-Toimela spectrum (\ref{ratethermal3}), and 
$a$, $b$, $n$ are adjustable parameters. 
The parameter $a$ quantifies the dependence of the suppression on $M_t$, according to Eq.~(\ref{reynoldsmt}).
The parameter $b$ quantifies the breaking of $M_t$ scaling due to pre-equilibrium dynamics. 
The functional form (\ref{fitformula}) guarantees that the deviations from ideal hydrodynamics are linear in $\eta/s$ in the limit $\eta/s\rightarrow 0$, as implied by the dimensional analysis in Sec.~\ref{s:dimensional}. 
The fact that this functional form gives a satisfactory fit of our numerical results for a wide range of values of $\eta/s$ is a clear indication that our dilepton spectrum converges smoothly to the McLerran-Toimela spectrum in the limit $\eta/s\to 0$. 
The parameter $n$ specifies the dependence of pre-equilibrium effects on the Reynolds number, in the non-linear regime where these effects are large. 
Note that the mass spectrum $dN^{l^{+}l^{-}}/dM$ obtained by integrating Eq.~(\ref{fitformula}) over $M_t$ is a  much better approximation of our numerical results than Eq.~(28) of \cite{Coquet:2021lca}. 

The parametrization (\ref{fitformula}) implies that the spectrum is proportional to $1/M$ in the limit of large $\eta/s$, in agreement with  Eq.~(\ref{2dspectrum}). 
However, the calculation leading to Eq.~(\ref{2dspectrum}) is not strictly equivalent to our quantitative calculation for the following reason. 
The hypothesis leading to Eq.~(\ref{2dspectrum}) is Eq.~(\ref{2dboltzmann}), namely, that the momentum distribution has zero width in $p_z$ and is exponential in $p_t$.  
In the quantitative calculation, the width of the $p_z$ distribution also goes to 0 in the limit of large $\eta/s$,  but it is not strictly exponential in $p_t$ (it is a Bessel function $K_0(p_t/T)$). 
Nevertheless, the dependence of the dilepton spectrum on $M$ ends up being essentially the same in both cases. 

\begin{figure}[ht]
\begin{center}
\includegraphics[width=1\linewidth]{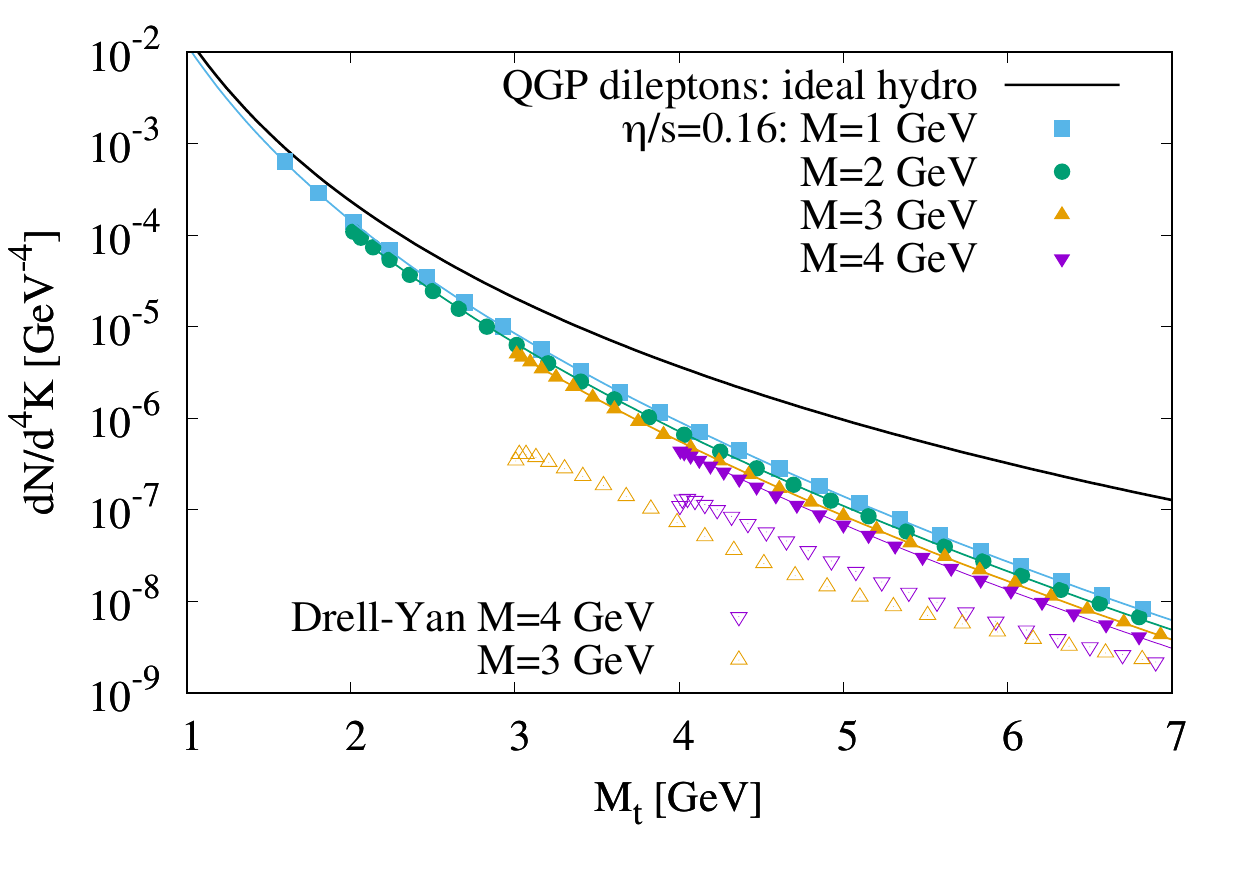} 
\includegraphics[width=1\linewidth]{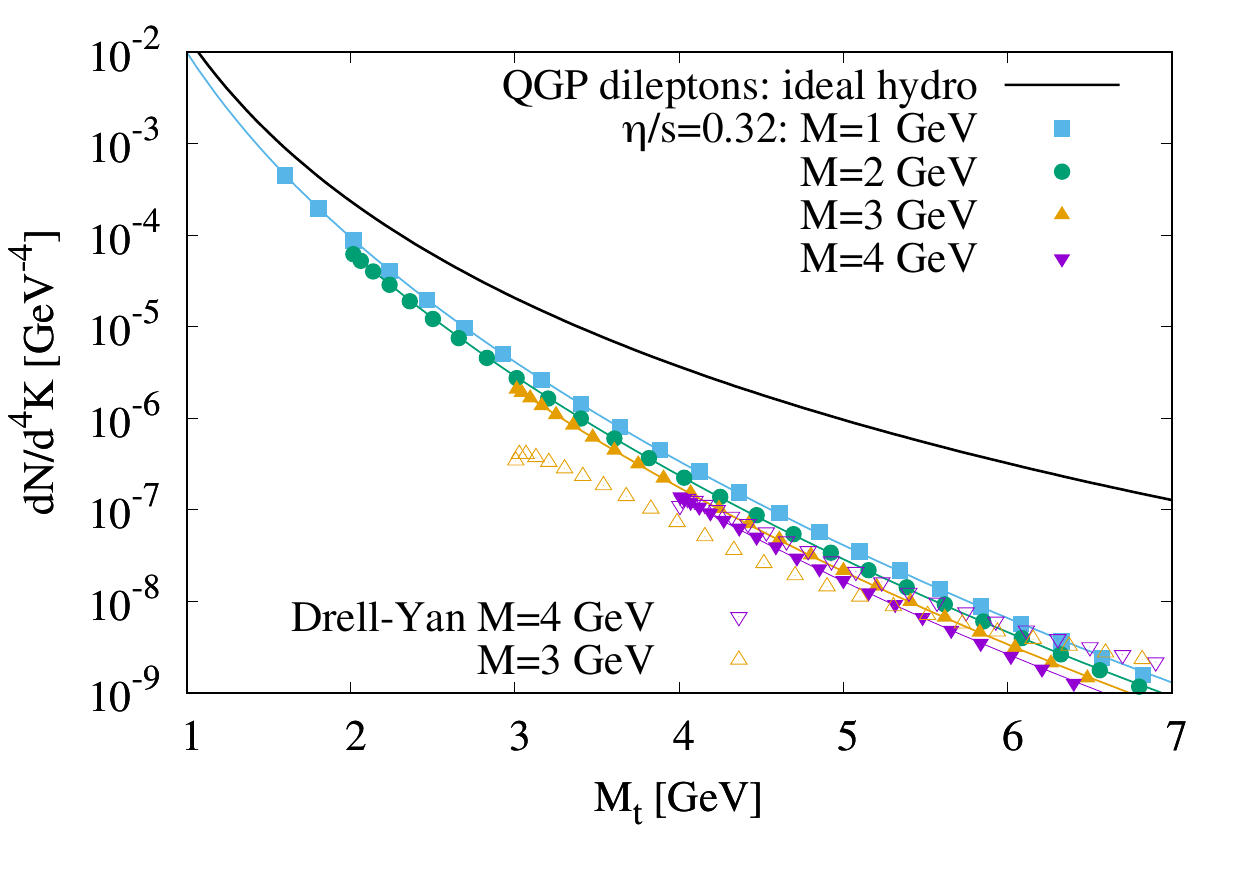} 
\end{center}
\caption{(Color online)
Full symbols: Expected dilepton invariant yield per event in $0-5\%$ central Pb+Pb collisions at $\sqrt{s_{\rm NN}}=5.02$~TeV, in the central rapidity window $|y|<1$, as a function of the transverse mass, for several values of the invariant mass $M$, and two values of the shear viscosity over entropy ratio $\eta/s$. 
The thick line is the McLerran-Toimela spectrum (\ref{ratethermal3}). 
The thin lines are the global fit of our results using Eq.~(\ref{fitformula}). 
Open symbols: Dilepton yield from the Drell-Yan process. We only plot the central value of the NLO+NLL calculation (see Sec.~\ref{s:backgrounds}). Uncertainties are shown in Fig.~\ref{fig:DY}.
}
\label{fig:allresults}
\end{figure}    

For each setup of our calculation, i.e., with or without quark suppression taken into account, we have carried out a global fit of all our results for $1.5<M_t<7$~GeV using Eq.~(\ref{fitformula}). 
The best-fit values with quark suppression  are $a=0.61$~GeV$^{-2}$, $b=1.6$~GeV$^{-2}$, $n=3.1$. 
Without quark suppression, they are $a=0.07$~GeV$^{-2}$, $b=2.4$~GeV$^{-2}$, $n=1.2$.\footnote{The error on $n$ is large for the results without quark suppression. 
If one fixes the value of $n$ to the same value as with quark suppression, the fit is almost as good, and the parameters $a$ and $b$ are not significantly modified, so that the smaller value of $n$ returned by the fit seems of little significance.} 
As expected from the discussion of Sec.~\ref{s:dimensional}, the breaking of $M_t$ scaling is larger without quark suppression, resulting in a larger value of $b$. 
The parameter $a$ is an order of magnitude smaller without quark suppression, which means that 90\% of the pre-equilibrium effects on the $M_t$ spectrum come from chemical equilibration. 
Note that from mere dimensional analysis, by comparing Eq.~(\ref{fitformula}) with Eq.~(\ref{reynoldsmt}), one expects $a\sim b\sim 1/(\tau T^3)\sim 2.5$~GeV$^{-2}$, where the numerical estimate is given by Eq.~(\ref{tauT3}). 
The values of $b$ returned by the fit are comparable, while those of $a$ are significantly smaller, in particular when quark suppression is not implemented. 

Fig.~\ref{fig:allresults} displays the dilepton yield per event for two values of the viscosity which roughly span the expected range in QCD~\cite{Christiansen:2014ypa}. 
By comparing with the McLerran-Toimela spectrum, one sees that pre-equilibrium dynamics suppresses dilepton production by at least a factor 10 for $M_t>5$~GeV.  
The dilepton yield is still mostly determined by $M_t$, and the breaking of $M_t$ scaling is a modest effect.

We now evaluate the robustness of our results with respect to transverse flow, which we have neglected. 
Transverse flow develops gradually over a time of the order of the nuclear radius. 
It becomes important for $\tau\gtrsim 5$~fm/c. 
If the fraction of the dileptons produced after $5$~fm/c is small, the dilepton yield is likely to have little sensitivity to transverse flow. 
We have calculated this fraction numerically and found that it only depends on $M_t$. 
It is roughly 25\% for $M_t=2$~GeV, but only 4\% for $M_t=3$~GeV. 
We conclude that for $M_t\lesssim 2$~GeV, sizable corrections from transverse flow are to be expected. 
At RHIC, it has been argued on the basis of simple dimensional arguments that these corrections are small in the intermediate mass region~\cite{Rapp:2014hha}. 
However, hydrodynamic calculations have shown that they are visible up to $M_t=2.5$~GeV~\cite{Deng:2010pq}.
Effects of transverse flow are larger at LHC than at RHIC. 
We intend to study them in a future publication. 

The dilepton spectrum is often characterized by its effective temperature $T_{\rm eff}$, defined as the inverse slope of the $M_t$ spectrum (Eq.~(\ref{teffideal})). 
It is interesting to note that the spectrum depends on $M$ only through a global factor in  Eq.~(\ref{fitformula}), so that $T_{\rm eff}$ still solely depends on $M_t$, as long as transverse flow can be neglected.\footnote{The rise and fall of $T_{\rm eff}$ as a function of $M$ observed by NA60~\cite{NA60:2008ctj,NA60:2008dcb} in the low-mass region can be ascribed to transverse flow.} 
In the limit of small $\eta/s$, Eq.~(\ref{fitformula}) gives: 
\begin{equation}
\label{teffviscous}
T_{\rm eff}(M_t)\simeq \frac{M_t}{6+2a\frac{\displaystyle\eta}{\displaystyle s} M_t^2}, 
\end{equation}
where $a\simeq 0.61$~GeV$^{-2}$. 
This equation shows that  pre-equilibrium dynamics decreases the effective temperature, and that the shear viscosity over entropy ratio at early times $\eta/s$ can be extracted from the inverse slope. 
Note that the inverse slope obtained in Ref.~\cite{Song:2018xca} using the Parton-Hadron String Dynamics (PHSD) model exceeds the McLerran-Toimela value (\ref{teffideal}) already at RHIC energies. 
We believe that this is due to the presence of hard particles of jets/mini-jets in the PHSD Monte Carlo simulations. 
This source of large invariant mass dileptons warrants further investigation. 

We now discuss the centrality and system-size dependence of QGP dilepton production. 
This dependence is encapsulated in the transverse area, $A_\perp$, and in the value of $\tau T^3$ at late times. 
Both quantities satisfy simple scaling laws as a function of the charged hadron multiplicity per unit pseudorapidity, $dN_{\rm ch}/d\eta \propto A_{\perp} \tau T^3$. 
The observation that the mean transverse momentum of hadrons $\langle p_t\rangle$ depends weakly on centrality and system size~\cite{ALICE:2018hza} implies that $A_\perp$ varies approximately like $(dN_{\rm ch}/d\eta)^{2/3}$~\cite{Gardim:2019brr} as a function of centrality and system size for fixed rapidity and collision energy. 
This in turn implies that $\tau T^3$, which is proportional to $(dN_{\rm ch}/d\eta)/A_\perp$~\cite{Coquet:2021lca}, scales like $(dN_{\rm ch}/d\eta)^{1/3}$. 
Eq.~(\ref{ratethermal3}) then shows that the McLerran-Toimela spectrum varies like $(dN_{\rm ch}/d\eta)^{4/3}$. 
In other words, the dilepton yield scales like the space-time volume, while the hadron yield scales like the volume at freeze-out.
The time component explains the extra factor  $(dN_{\rm ch}/d\eta)^{1/3}$.

Eq.~(\ref{reynoldsmt}) then shows that the relative modification of the dilepton yield due to pre-equilibrium dynamics varies with system size and centrality like $(dN_{\rm ch}/d\eta)^{-1/3}$. 
This implies that the parameters $a$ and $b$ in Eq.~(\ref{fitformula}) are also proportional to $(dN_{\rm ch}/d\eta)^{-1/3}$.
However, note that local event-to-event fluctuations of the initial density~\cite{Aguiar:2001ac}, which we neglect, will break this simple scaling. 

Similar dimensional arguments can be used to predict the dependence of QGP dilepton production on the collision energy $\sqrt{s_{\rm NN}}$. 
For a given collision system, the transverse area $A_\perp$ is approximately independent of $\sqrt{s_{\rm NN}}$, while the hadron multiplicity $dN_{\rm ch}/d\eta$ increases with $\sqrt{s_{\rm NN}}$~\cite{ALICE:2015juo}. 
Therefore, $\tau T^3$ scales with energy like $dN_{\rm ch}/d\eta$. 
This implies that the McLerran-Toimela spectrum (\ref{ratethermal3}) is proportional to $(dN_{\rm ch}/d\eta)^2$. 
On the other hand, the coefficients $a$ and $b$, which govern the modifications due to pre-equilibrium effects, are proportional to the inverse Reynolds number (\ref{reynoldsmt}), i.e., to $(dN_{\rm ch}/d\eta)^{-1}$. 

Finally, the dependence on rapidity $y$ follows the same scaling rules as the dependence on collision energy, up to the replacement of $dN_{\rm ch}/d\eta$ with $dN_{\rm ch}/dy$.\footnote{Our calculation setup assumes longitudinal boost invariance, but the results can still be applied if the multiplicity depends on rapidity, since the  longitudinal pressure gradient has a negligible effect at LHC energy~\cite{Ollitrault:2007du}.} 
The McLerran-Toimela spectrum is proportional to $(dN_{\rm ch}/dy)^2$, while $a$ and $b$ are proportional to $(dN_{\rm ch}/dy)^{-1}$. 
The dilepton spectrum is therefore maximum at mid-rapidity, where $dN_{\rm ch}/dy$ is maximum~\cite{ALICE:2016fbt}. 

\section{Background from the Drell-Yan process}
\label{s:backgrounds}

\begin{figure}[ht]
\begin{center}
\includegraphics[width=\linewidth]{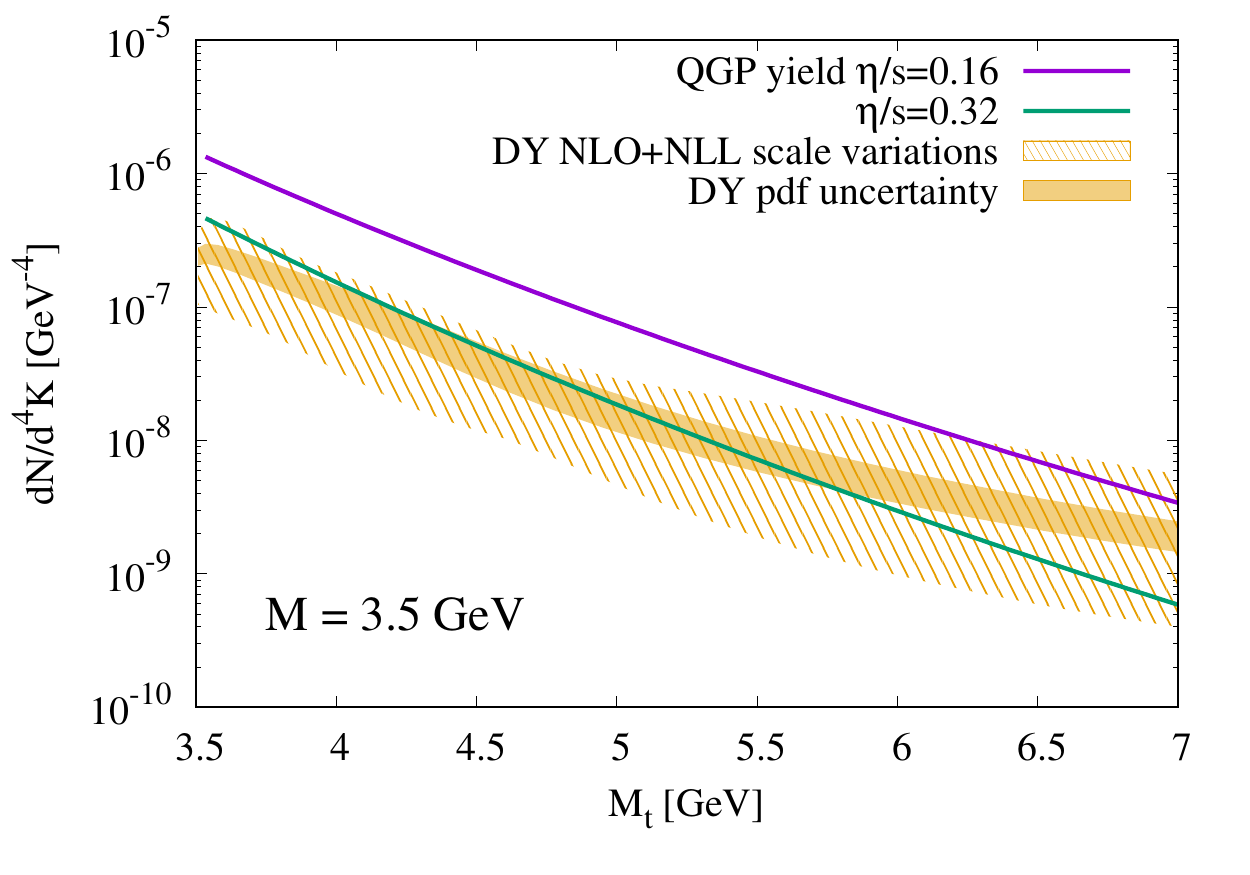} 
\includegraphics[width=\linewidth]{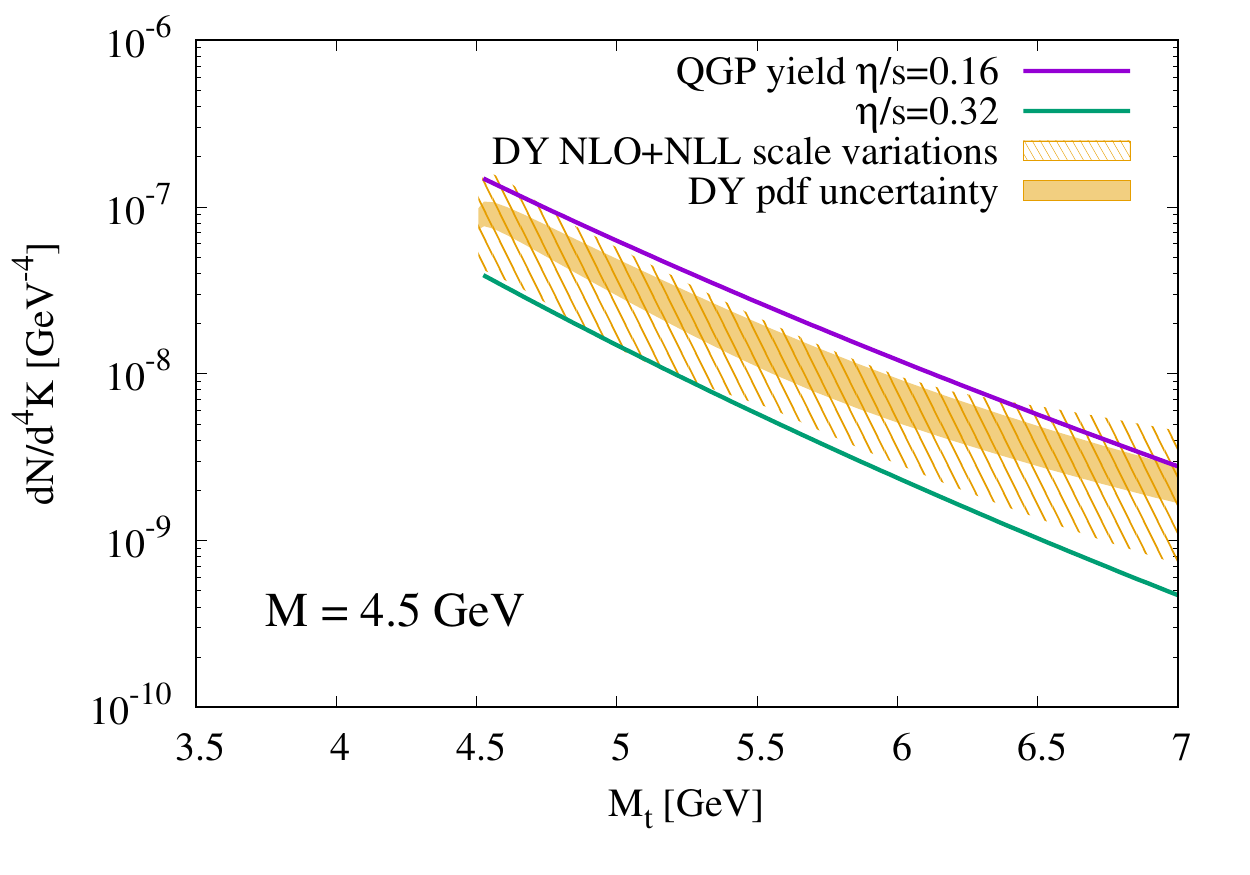} 
\end{center}
\caption{(Color online)
Comparison between the dilepton yield from  production in the QGP and from the Drell-Yan process for two fixed values of the invariant mass $M=3.5$~GeV (top) and $M=4.5$~GeV (bottom). 
Lines: QGP production with pre-equilibrium dynamics and quark suppression taken into account, for $\eta/s=0.16$ (upper line) and $\eta/s=0.32$ (lower line). 
Dark shaded band: Drell Yan process, with  uncertainty from the parton distribution function. 
Light shaded band: uncertainty on Drell-Yan from the renormalization and factorization scales. 
Both bands are obtained by taking the envelope of the results obtained by varying the model parameters. 
}
\label{fig:DY}
\end{figure}    

The main backgrounds to QGP dilepton production in the intermediate mass region, besides the large $J/\psi$ peak around $M\simeq 3.1$~GeV, are semileptonic decays of heavy quark hadrons, and Drell-Yan production in the initial state. 
The number of dileptons from charm hadron decays is expected to exceed the QGP dilepton yield for $\sqrt{s_{\rm NN}}>40$~GeV~\cite{Song:2018xca}. 
Charm decays are indeed observed to be the dominant source of dileptons at LHC energies~\cite{ALICE:2018ael}. This source of background can be rejected based on the finite lifetime of the heavy quark hadrons since the leptons from this source do not originate from the primary vertex. Despite the small lifetime of charm hadron ground states of $c \tau \approx$  50-410~$\mu$m, the partial rejection of these decay leptons in heavy-ion collisions is feasible and will strongly improve with the future detector projects LHCb Upgrade 2~\cite{LHCb:2018roe} and ALICE 3~\cite{Adamova:2019vkf}. We consider in this phenomenological publication only the irreducible background from the Drell-Yan process.

We compute the production of dileptons by the Drell-Yan process using the Drell-Yan Turbo software package~\cite{Camarda:2019zyx}. 
The cross section is evaluated at Next-to-Leading Order~(NLO) in the strong coupling constant $\alpha_s$. 
In addition, the calculation employs a resummation at small transverse momentum at next-to-leading logarithm~(NLL). 
For the non-perturbative contribution to the form factor, the same Gaussian form is chosen as in~\cite{Camarda:2019zyx}. 
We evaluate the uncertainty on the Drell Yan spectrum in the following way: 
We vary the renormalization and factorization scales by a factor two independently resulting in 8 variations with respect to the default choice. We take into account the uncertainty on the parton distribution function in the EPPS parametrization~\cite{Eskola:2016oht}.

The central value of our calculation is plotted in Fig.~\ref{fig:allresults} for two values of the invariant mass $M$. 
The slope of the Drell-Yan $M_t$ spectrum  is roughly similar to that of the QGP spectrum, albeit slightly flatter.  
The main difference between the two spectra is  the normalization, which depends on the invariant mass $M$. 
Drell-Yan production is enhanced for larger values of $M$ at a given $M_t$, contrary to QGP production. 
The physical explanation is that the momenta of the incoming quark and antiquark responsible for Drell-Yan production are mostly longitudinal, so that smaller values of the transverse momentum $k_t$ (corresponding to larger values of $M$ at a given $M_t$) are preferred. 
The kinematics of early QGP production is opposite, in the sense that longitudinal momenta in the QGP are typically smaller than  transverse momenta. 
This means that the breaking of $M_t$ scaling alone provide a handle to distinguish between QGP and Drell-Yan production. 

Drell-Yan gradually takes over QGP production as $M$ increases. 
In order to evaluate where the transition occurs, we compare both spectra in Fig.~\ref{fig:DY} for two values of $M$ above the $J/\psi$ peak, and for the two values of $\eta/s$ used in Fig.~\ref{fig:allresults}. 
The uncertainty on the Drell-Yan spectrum from the renormalization and factorization scales is displayed as a light shaded band.
The uncertainty from parton distribution function is displayed as a dark shaded band. 
One sees that the scale is the dominant source of uncertainty. 
The top panel shows that QGP production is likely to dominate over Drell-Yan for $M$ up to $3.5$~GeV, at least for the lowest values of $M_t$. 
Looking at the bottom panel, it seems unlikely that QGP dileptons can be isolated above $M=4.5$~GeV. 
The precise value of $M$ above which Drell-Yan dominates over QGP production depends on the value of $\eta/s$ at early times. 
Drell-Yan calculations can be also performed down to lower invariant masses. However, at low invariant masses the pQCD calculation exhibits increasingly large scale uncertainties. Furthermore, the non-perturbative set-up of the used code (NLO+NLL) was developed based on high invariant masses (Z-mass) and we hence do not want to stretch into a regime where the chosen approach may not be the most appropriate one that one could apply. Nevertheless, based on our estimates, the production from the QGP and the preequilibrium phase is expected to outshine the Drell-Yan contribution at smaller invariant masses.

The calculations displayed in the figures are carried out at mid-rapidity. 
The rapidity dependence of the Drell-Yan is milder than that of the QGP spectrum and goes in the opposite direction: 
Its minimum is at midrapidity. 
On the other hand, the background from semileptonic decays of heavy quarks is easier to eliminate at larger rapidities because the secondary vertex is farther from the collision point. 

\section{Conclusions}
\label{s:conclusions}

We have calculated the spectrum of dileptons  produced by the quark-gluon plasma in ultrarelativistic heavy-ion collisions for invariant masses larger than $1.2$~GeV, with input from  state-of-the-art QCD kinetic theory to model the kinetic and chemical equilibration of the QGP at early times.  
The invariant spectrum $dN^{l^+l^-}/d^4K$ depends mostly on the transverse mass $M_t$. 
The underpopulation of quarks at early times results in a steeper $M_t$ spectrum. 
The viscosity over entropy ratio, which determines the equilibration time of the QGP, can be inferred by measuring the slope of the spectrum. 

The anisotropy of the momentum distribution at early times breaks transverse mass scaling, by suppressing the production of higher invariant masses $M$. 
Interestingly, the trend is opposite for dileptons produced by the Drell-Yan process, which is enhanced for larger $M$. 
Therefore, one can distinguish experimentally QGP production from Drell-Yan production by studying the variation of the dilepton yield as a function of $M$ at fixed $M_t$. 

We have introduced a simple parametrization (\ref{fitformula}), from which one can infer the dependence of QGP dilepton production on centrality, system size, rapidity, and collision energy. 
Our modelization can be improved in several ways.
For the larger values of $M_t$, the contribution of minijets~\cite{Song:2018xca,Paatelainen:2013eea} warrants additional study. 
For the smaller values of $M_t$, one should take transverse flow into account, as its effect on dilepton production is likely to be significant for transverse masses $M_t\lesssim 2$~GeV. 
It has been studied in detail in the low-mass region~\cite{Vujanovic:2013jpa}. 
In the intermediate mass region, detailed studies of transverse flow have been carried out~\cite{Ryblewski:2015hea,Kasmaei:2018oag}. 
However, the effect of transverse flow on the slope on the $M_t$ spectrum has only been studied at RHIC energy~\cite{Deng:2010pq}, and deserves further studies at higher energies.  

\section*{Acknowledgments} 
This work is supported in part by the Deutsche Forschungsgemeinschaft (DFG, German Research Foundation) through the CRC-TR 211 ``Strong-interaction matter under extreme conditions'' project number 315477589 – TRR 211 and in part in the framework of the GLUODYNAMICS project funded by the ``P2IO LabEx (ANR-10-LABX-0038)'' in the framework ``Investissements d'Avenir'' (ANR-11-IDEX-0003-01) managed by the Agence Nationale de la Recherche (ANR), France.


\begin{thebibliography}{99}

\bibitem{Busza:2018rrf}
W.~Busza, K.~Rajagopal and W.~van der Schee,
Ann. Rev. Nucl. Part. Sci. \textbf{68} (2018), 339-376
doi:10.1146/annurev-nucl-101917-020852
[arXiv:1802.04801 [hep-ph]].

\bibitem{Martinez:2008di}
M.~Martinez and M.~Strickland,
Phys. Rev. C \textbf{78} (2008), 034917
doi:10.1103/PhysRevC.78.034917
[arXiv:0805.4552 [hep-ph]].

\bibitem{Rapp:2013nxa}
R.~Rapp,
Adv. High Energy Phys. \textbf{2013} (2013), 148253
doi:10.1155/2013/148253
[arXiv:1304.2309 [hep-ph]].

\bibitem{Song:2018xca}
T.~Song, W.~Cassing, P.~Moreau and E.~Bratkovskaya,
Phys. Rev. C \textbf{97} (2018) no.6, 064907
doi:10.1103/PhysRevC.97.064907
[arXiv:1803.02698 [nucl-th]].

\bibitem{McLerran:1984ay}
L.~D.~McLerran and T.~Toimela,
Phys. Rev. D \textbf{31} (1985), 545
doi:10.1103/PhysRevD.31.545

\bibitem{Coquet:2021lca}
M.~Coquet, X.~Du, J.~Y.~Ollitrault, S.~Schlichting and M.~Winn,
Phys. Lett. B \textbf{821} (2021), 136626
doi:10.1016/j.physletb.2021.136626
[arXiv:2104.07622 [nucl-th]].

\bibitem{Drell:1970wh}
S.~D.~Drell and T.~M.~Yan,
Phys. Rev. Lett. \textbf{25} (1970), 316-320
[erratum: Phys. Rev. Lett. \textbf{25} (1970), 902]
doi:10.1103/PhysRevLett.25.316

\bibitem{Hagedorn:1965st}
R.~Hagedorn,
Nuovo Cim. Suppl. \textbf{3} (1965), 147-186
CERN-TH-520.

\bibitem{Aachen-Berlin-Bonn-CERN-Cracow-Heidelberg-Warsaw:1976gom}
J.~Bartke \textit{et al.} [Aachen-Berlin-Bonn-CERN-Cracow-Heidelberg-Warsaw],
Nucl. Phys. B \textbf{120} (1977), 14-22
doi:10.1016/0550-3213(77)90092-X

\bibitem{Becattini:1997rv}
F.~Becattini and U.~W.~Heinz,
Z. Phys. C \textbf{76} (1997), 269-286
[erratum: Z. Phys. C \textbf{76} (1997), 578]
doi:10.1007/s002880050551
[arXiv:hep-ph/9702274 [hep-ph]].

\bibitem{Altenkamper:2017qot}
L.~Altenk\"amper, F.~Bock, C.~Loizides and N.~Schmidt,
Phys. Rev. C \textbf{96} (2017) no.6, 064907
doi:10.1103/PhysRevC.96.064907
[arXiv:1710.01933 [hep-ph]].

\bibitem{Bjorken:1982qr}
J.~D.~Bjorken,
Phys. Rev. D \textbf{27} (1983), 140-151
doi:10.1103/PhysRevD.27.140

\bibitem{Ghiglieri:2014kma}
J.~Ghiglieri and G.~D.~Moore,
JHEP \textbf{12} (2014), 029
doi:10.1007/JHEP12(2014)029
[arXiv:1410.4203 [hep-ph]].

\bibitem{Ding:2016hua}
H.~T.~Ding, O.~Kaczmarek and F.~Meyer,
Phys. Rev. D \textbf{94} (2016) no.3, 034504
doi:10.1103/PhysRevD.94.034504
[arXiv:1604.06712 [hep-lat]].

\bibitem{Jackson:2019yao}
G.~Jackson and M.~Laine,
JHEP \textbf{11} (2019), 144
doi:10.1007/JHEP11(2019)144
[arXiv:1910.09567 [hep-ph]].

\bibitem{Laine:2015iia}
M.~Laine,
PoS \textbf{CPOD2014} (2015), 065
doi:10.22323/1.217.0065
[arXiv:1502.05796 [hep-ph]].

\bibitem{Baier:2007ix}
R.~Baier, P.~Romatschke, D.~T.~Son, A.~O.~Starinets and M.~A.~Stephanov,
JHEP \textbf{04} (2008), 100
doi:10.1088/1126-6708/2008/04/100
[arXiv:0712.2451 [hep-th]].

\bibitem{NA60:2008ctj}
R.~Arnaldi \textit{et al.} [NA60],
Eur. Phys. J. C \textbf{61} (2009), 711-720
doi:10.1140/epjc/s10052-009-0878-5
[arXiv:0812.3053 [nucl-ex]].

\bibitem{Rapp:2014hha}
R.~Rapp and H.~van Hees,
Phys. Lett. B \textbf{753} (2016), 586-590
doi:10.1016/j.physletb.2015.12.065
[arXiv:1411.4612 [hep-ph]].

\bibitem{Tripolt:2020dac}
R.~A.~Tripolt,
Nucl. Phys. A \textbf{1005} (2021), 121755
doi:10.1016/j.nuclphysa.2020.121755
[arXiv:2001.11232 [hep-ph]].

\bibitem{HADES:2019auv}
J.~Adamczewski-Musch \textit{et al.} [HADES],
Nature Phys. \textbf{15} (2019) no.10, 1040-1045
doi:10.1038/s41567-019-0583-8

\bibitem{Kajantie:1986cu}
K.~Kajantie, M.~Kataja, L.~D.~McLerran and P.~V.~Ruuskanen,
Phys. Rev. D \textbf{34} (1986), 811
doi:10.1103/PhysRevD.34.811

\bibitem{Ollitrault:2007du}
J.~Y.~Ollitrault,
Eur. J. Phys. \textbf{29} (2008), 275-302
doi:10.1088/0143-0807/29/2/010
[arXiv:0708.2433 [nucl-th]].

\bibitem{Shuryak:1992wc}
E.~V.~Shuryak,
Phys. Rev. Lett. \textbf{68} (1992), 3270-3272
doi:10.1103/PhysRevLett.68.3270

\bibitem{Kurkela:2018oqw}
A.~Kurkela and A.~Mazeliauskas,
Phys. Rev. D \textbf{99} (2019) no.5, 054018
doi:10.1103/PhysRevD.99.054018
[arXiv:1811.03068 [hep-ph]].

\bibitem{Shuryak:1992bt}
E.~V.~Shuryak and L.~Xiong,
Phys. Rev. Lett. \textbf{70} (1993), 2241-2244
doi:10.1103/PhysRevLett.70.2241
[arXiv:hep-ph/9301218 [hep-ph]].

\bibitem{Vredevoogd:2008id}
J.~Vredevoogd and S.~Pratt,
Phys. Rev. C \textbf{79} (2009), 044915
doi:10.1103/PhysRevC.79.044915
[arXiv:0810.4325 [nucl-th]].

\bibitem{Kurkela:2018wud}
A.~Kurkela, A.~Mazeliauskas, J.~F.~Paquet, S.~Schlichting and D.~Teaney,
Phys. Rev. Lett. \textbf{122} (2019) no.12, 122302
doi:10.1103/PhysRevLett.122.122302
[arXiv:1805.01604 [hep-ph]].

\bibitem{Deng:2010pq}
J.~Deng, Q.~Wang, N.~Xu and P.~Zhuang,
Phys. Lett. B \textbf{701} (2011), 581-586
doi:10.1016/j.physletb.2011.06.027
[arXiv:1009.3091 [nucl-th]].

\bibitem{NA60:2007lzy}
R.~Arnaldi \textit{et al.} [NA60],
Phys. Rev. Lett. \textbf{100} (2008), 022302
doi:10.1103/PhysRevLett.100.022302
[arXiv:0711.1816 [nucl-ex]].

\bibitem{Paatelainen:2013eea}
R.~Paatelainen, K.~J.~Eskola, H.~Niemi and K.~Tuominen,
Phys. Lett. B \textbf{731} (2014), 126-130
doi:10.1016/j.physletb.2014.02.018
[arXiv:1310.3105 [hep-ph]].

\bibitem{Churchill:2020uvk}
J.~Churchill, L.~Yan, S.~Jeon and C.~Gale,
Phys. Rev. C \textbf{103} (2021) no.2, 024904
doi:10.1103/PhysRevC.103.024904
[arXiv:2008.02902 [hep-ph]].

\bibitem{Du:2020dvp}
X.~Du and S.~Schlichting,
Phys. Rev. D \textbf{104} (2021) no.5, 054011
doi:10.1103/PhysRevD.104.054011
[arXiv:2012.09079 [hep-ph]].

\bibitem{Heller:2015dha}
M.~P.~Heller and M.~Spalinski,
Phys. Rev. Lett. \textbf{115} (2015) no.7, 072501
doi:10.1103/PhysRevLett.115.072501
[arXiv:1503.07514 [hep-th]].

\bibitem{Heller:2016rtz}
M.~P.~Heller, A.~Kurkela, M.~Spali\'nski and V.~Svensson,
Phys. Rev. D \textbf{97} (2018) no.9, 091503
doi:10.1103/PhysRevD.97.091503
[arXiv:1609.04803 [nucl-th]].

\bibitem{Giacalone:2019ldn}
G.~Giacalone, A.~Mazeliauskas and S.~Schlichting,
Phys. Rev. Lett. \textbf{123} (2019) no.26, 262301
doi:10.1103/PhysRevLett.123.262301
[arXiv:1908.02866 [hep-ph]].

\bibitem{Christiansen:2014ypa}
N.~Christiansen, M.~Haas, J.~M.~Pawlowski and N.~Strodthoff,
Phys. Rev. Lett. \textbf{115} (2015) no.11, 112002
doi:10.1103/PhysRevLett.115.112002
[arXiv:1411.7986 [hep-ph]].

\bibitem{Policastro:2001yc}
G.~Policastro, D.~T.~Son and A.~O.~Starinets,
Phys. Rev. Lett. \textbf{87} (2001), 081601
doi:10.1103/PhysRevLett.87.081601
[arXiv:hep-th/0104066 [hep-th]].

\bibitem{NA60:2008dcb}
R.~Arnaldi \textit{et al.} [NA60],
Eur. Phys. J. C \textbf{59} (2009), 607-623
doi:10.1140/epjc/s10052-008-0857-2
[arXiv:0810.3204 [nucl-ex]].

\bibitem{ALICE:2018hza}
S.~Acharya \textit{et al.} [ALICE],
Phys. Lett. B \textbf{788} (2019), 166-179
doi:10.1016/j.physletb.2018.10.052
[arXiv:1805.04399 [nucl-ex]].

\bibitem{Gardim:2019brr}
F.~G.~Gardim, G.~Giacalone and J.~Y.~Ollitrault,
Phys. Lett. B \textbf{809} (2020), 135749
doi:10.1016/j.physletb.2020.135749
[arXiv:1909.11609 [nucl-th]].

\bibitem{Aguiar:2001ac}
C.~E.~Aguiar, Y.~Hama, T.~Kodama and T.~Osada,
Nucl. Phys. A \textbf{698} (2002), 639-642
doi:10.1016/S0375-9474(01)01447-6
[arXiv:hep-ph/0106266 [hep-ph]].

\bibitem{ALICE:2015juo}
J.~Adam \textit{et al.} [ALICE],
Phys. Rev. Lett. \textbf{116} (2016) no.22, 222302
doi:10.1103/PhysRevLett.116.222302
[arXiv:1512.06104 [nucl-ex]].

\bibitem{ALICE:2016fbt}
J.~Adam \textit{et al.} [ALICE],
Phys. Lett. B \textbf{772} (2017), 567-577
doi:10.1016/j.physletb.2017.07.017
[arXiv:1612.08966 [nucl-ex]].

\bibitem{ALICE:2018ael}
S.~Acharya \textit{et al.} [ALICE],
Phys. Rev. C \textbf{99} (2019) no.2, 024002
doi:10.1103/PhysRevC.99.024002
[arXiv:1807.00923 [nucl-ex]].

\bibitem{LHCb:2018roe}
R.~Aaij \textit{et al.} [LHCb],
[arXiv:1808.08865 [hep-ex]].

\bibitem{Adamova:2019vkf}
D.~Adamov\'a, G.~Aglieri Rinella, M.~Agnello, Z.~Ahammed, D.~Aleksandrov, A.~Alici, A.~Alkin, T.~Alt, I.~Altsybeev and D.~Andreou, \textit{et al.}
[arXiv:1902.01211 [physics.ins-det]].

\bibitem{Camarda:2019zyx}
S.~Camarda, M.~Boonekamp, G.~Bozzi, S.~Catani, L.~Cieri, J.~Cuth, G.~Ferrera, D.~de Florian, A.~Glazov and M.~Grazzini, \textit{et al.}
Eur. Phys. J. C \textbf{80} (2020) no.3, 251
[erratum: Eur. Phys. J. C \textbf{80} (2020) no.5, 440]
doi:10.1140/epjc/s10052-020-7757-5
[arXiv:1910.07049 [hep-ph]].

\bibitem{Eskola:2016oht}
K.~J.~Eskola, P.~Paakkinen, H.~Paukkunen and C.~A.~Salgado,
Eur. Phys. J. C \textbf{77} (2017) no.3, 163
doi:10.1140/epjc/s10052-017-4725-9
[arXiv:1612.05741 [hep-ph]].

\bibitem{Vujanovic:2013jpa}
G.~Vujanovic, C.~Young, B.~Schenke, R.~Rapp, S.~Jeon and C.~Gale,
Phys. Rev. C \textbf{89} (2014) no.3, 034904
doi:10.1103/PhysRevC.89.034904
[arXiv:1312.0676 [nucl-th]].

\bibitem{Ryblewski:2015hea}
R.~Ryblewski and M.~Strickland,
Phys. Rev. D \textbf{92} (2015) no.2, 025026
doi:10.1103/PhysRevD.92.025026
[arXiv:1501.03418 [nucl-th]].

\bibitem{Kasmaei:2018oag}
B.~S.~Kasmaei and M.~Strickland,
Phys. Rev. D \textbf{99} (2019) no.3, 034015
doi:10.1103/PhysRevD.99.034015
[arXiv:1811.07486 [hep-ph]].

\end{thebibliography}
\end{document}